\documentclass[conference]{IEEEtran}
\IEEEoverridecommandlockouts
\usepackage{comment}
\usepackage{lipsum}
\usepackage{epsfig,graphics,amssymb}
\usepackage{amscd}
\usepackage{amsmath}
\usepackage{enumerate}
\usepackage{theorem}
\usepackage{cite}
\usepackage{stfloats}
\usepackage{epsfig}
\usepackage{verbatim}
\usepackage{color}
\theoremstyle{plain} \theorembodyfont{\itshape}
\theoremheaderfont{\scshape}
\newtheorem{theorem}{Theorem}
\theorembodyfont{\itshape} \theoremheaderfont{\scshape}

\theoremstyle{plain} \theorembodyfont{\itshape}
\theoremheaderfont{\scshape}
\newtheorem{corollary}{Corollary}

\title{S-AMP: Approximate Message Passing for\\\quad General Matrix Ensembles}
\author{\IEEEauthorblockA{Burak \c{C}akmak \\ Department of Electronic Systems \\ Aalborg University \\ 9220 Aalborg, Denmark\\Email: buc@es.aau.dk} \and  
\IEEEauthorblockA{Ole Winther\\
DTU Compute\\
Technical University of Denmark\\2800 Lyngby, Denmark\\Email: olwi@dtu.dk}  
\and \IEEEauthorblockA{Bernard H. Fleury \\ Department of Electronic Systems \\ Aalborg University \\ 9220 Aalborg, Denmark\\Email: fleury@es.aau.dk}}
\begin{document}
\def\mathlette#1#2{{\mathchoice{\mbox{#1$\displaystyle #2$}}%
                               {\mbox{#1$\textstyle #2$}}%
                               {\mbox{#1$\scriptstyle #2$}}%
                               {\mbox{#1$\scriptscriptstyle #2$}}}}
\newcommand{\matr}[1]{\mathlette{\boldmath}{#1}}
\newcommand{\RR}{\mathbb{R}}
\newcommand{\CC}{\mathbb{C}}
\newcommand{\NN}{\mathbb{N}}
\newcommand{\ZZ}{\mathbb{Z}}
\maketitle

\begin{abstract}
In this work we propose a novel iterative estimation algorithm for linear observation systems called S-AMP whose fixed points are the stationary points of the exact Gibbs free energy under a set of (first- and second-) moment consistency constraints in the large system limit. S-AMP extends the approximate message-passing (AMP) algorithm to general matrix ensembles. The generalization is based on the S-transform (in free probability) of the spectrum of the measurement matrix. Furthermore, we show that the optimality of S-AMP follows directly from its design rather than from solving a separate optimization problem as done for AMP.
\end{abstract}


\section{Introduction}
Consider an $N\times K$ linear observation model described by
\begin{equation}
\matr y=\matr A\matr x +\matr w \label{system}
\end{equation}
where $\matr A\in\mathbb R^{N\times K}$, $\matr{x}\in \mathbb R^{K\times 1}$, $\matr y\in \mathbb R^{N\times 1}$, and $\matr w\in\mathbb R^{N\times 1}$ are the measurement matrix, the vector to be recovered, the measurement vector, and a white Gaussian noise vector, respectively. The entries of $\matr w$ have variance $\sigma_w^2$. In \cite{Donoha1} the authors propose a recovery scheme for $\matr x$, given $\matr A$ and $\matr y$, called Approximate Message Passing (AMP) algorithm, which starting from an initial guess $\matr \mu^{0}=\matr 0$, proceeds iteratively  according to
\begin{eqnarray}
\matr \mu^{t+1}&=&\eta_{t}\left(\matr A^\dagger\matr z^{t}+\matr\mu^t\right)\label{AMP1}\\
\matr z^t &=& \matr y-\matr A\matr \mu^t+\frac{1}{\alpha} \left<\eta_{t-1}'(\matr A^\dagger\matr z^{t-1}+\matr\mu^{t-1}) \right>\matr z^{t-1}. \label{AMP2}
\end{eqnarray}
The scalar functions $\eta_t$, $t\geq 0$, in \eqref{AMP1} are obtained by applying an additional optimization procedure based upon the so-called state evolution formula for the underlying measurement matrix ensemble \cite{Bayati}. In \eqref{AMP2}, $\eta'_t(x)=d\eta_t(x)/dx$, $t\geq 0$. 
Moreover for a vector $\matr u\triangleq(u_1,\dots,u_K)$, $\left<\matr u\right>\triangleq\sum_{k=1}^{K} u_k/K$ and $\alpha\triangleq N/K$. The vectors $\matr \mu^t$ and $\matr z^t$ are referred to as the current estimate of $\matr x$ and the corresponding residual, respectively. Finally $(\cdot)^\dagger$ denotes transposition.

AMP has two appealing properties. Firstly, when the entries of $\matr A$ are independent identically distributed (iid) Gaussian with zero mean and variance $1/N$, AMP yields the minimum mean square error (MMSE) estimator in the large system limit \cite{Bayati}. Secondly, AMP includes a so-called Onsager reaction term, i.e, $\alpha^{-1} \left<\eta_{t-1}'(\cdot)\right>\matr z^{t-1}$ in \eqref{AMP2}, that corrects the naive mean field approximation. In statistical physics such a technique is known as the Thouless-Anderson-Palmer (TAP) correction \cite{Manfred}. 

The adaptive TAP (ADATAP) mean field theory was introduced in \cite{Ole}. In ADATAP the form of Onsager reaction term depends on the measurement matrix, see \cite[Eq.~(20) \& (51)]{Ole}. Indeed, a connection between ADATAP and AMP has been recently realized in \cite{Kab}. The connection is based on some approximations of the Gibbs free energy, which are derived using the replica method, see \cite[Eq.~(10) \& (11)]{Kab} and the references therein. 

Inference techniques based on the free energy optimization have become popular in the literature of information theory \cite{Yedida,Erwin} and in machine learning  
\cite{Heskes},\cite{Ole5} and references therein. The important results exploited in this contribution is that the fixed points of belief propagation (BP) and expectation propagation (EP) are the stationary points of the Bethe Free energy (BFE) under a set of marginalization consistency constraints \cite{Yedida} and moment consistency constraints \cite{Heskes}, respectively.

The conventional approximate message passing methods presented in the literature are based on a Gaussian approximation of loopy BP on a dense graph, \cite{Donoha2,Rangan,Florent}. By  contrast, the  method presented in this paper is based on probabilistic inference on a tree graph. Specifically we consider an exact Gibbs free energy formulation (i.e. a BFE formulation on a tree probabilistic graph) under first and second-moment consistency constraints. Our analysis relies on the stationary point equations of the constrained Gibbs free energy. In particular we propose a novel algorithm whose fixed points are the stationary points of the constrained Gibbs free energy in the large system limit. This algorithm -- we coin it S-AMP -- executes the following iteration steps:
\begin{eqnarray}
\matr \mu^{t+1}&=& \eta_{t}\left(\matr A^\dagger\matr z^{t}+\matr\mu^t \right) \label{keyb}\\
\matr z^t &=& \matr y-\matr A\matr \mu^t+\left(1-\frac{1}{{\rm s}_{\matr A}^{t-1}}\right)\matr z^{t-1}\\
{\rm s}_{\matr A}^{t-1}&\triangleq & {\rm S}_{\matr A}\left(-\left <\eta_{t-1}'(\matr A^\dagger\matr z^{t-1}+\matr\mu^{t-1})\right>\right)\label{key}
\end{eqnarray}
with $\rm S_{\matr A}$ denoting the S-transform of the asymptotic eigenvalue distribution (AED) of $\matr A^\dagger \matr A$ (see, e.g. \cite{tulino}). Later in the paper we will show that the optimality of S-AMP follows by its design rather than based upon an optimization procedure as in \cite{Bayati}.     

To show that AMP is a special case of S-AMP, let the entries of $\matr A$ be iid with zero mean variance $1/N$. Then, as $N,K\to \infty$ with the ratio $\alpha=N/K$ fixed, $\rm S_{\matr A}(\omega)=1/(1+\omega/\alpha)$ \cite[Eq. (2.87)]{tulino}. Inserting this expression in \eqref{key} we obtain the iteration steps \eqref{AMP1}-\eqref{AMP2} of AMP. 

\emph{Notation:} The entries of the $N\times K$ matrix $\matr X$ are denoted by $X_{nk}$, $n\in \mathcal N$ and $k\in \mathcal K$ with $\mathcal N\triangleq \{1,\dots, N\}$ and $\mathcal K\triangleq\{1,\dots, K\}$. The entries of a vector $\matr u\in \mathbb R^{K\times 1}$ are indicated by $u_k$. The Gaussian probability density function (pdf) is denoted by $N(\cdot \vert \matr \mu, \matr\Sigma)$ with mean $\matr \mu$ and the covariance $\matr \Sigma$. Throughout the paper we assume that $\matr A^\dagger \matr A$ has almost surely an AED as $N,K\to \infty$ with the ratio $\alpha=N/K$ fixed.
\section{Gibbs Free Energy with Moment Constraints}
Consider the $N\times K$ linear observation model \eqref{system}. For Bayesian inference, we assign a prior $p_k(x_k)$ for all $k\in \mathcal K$. Hence the joint posterior pdf can be written as
\begin{equation}
\vspace{-0.05cm}
p(\matr x\vert \matr y)=\frac{1}{Z}p(\matr y\vert \matr x)\prod_{k\in \mathcal { K}}p_k(x_k) \label{fac} 
\end{equation}
with $p(\matr y\vert\matr x)$ and $Z$ denoting the likelihood given by \eqref{system} and a normalization constant, respectively. The factor graph representation of \eqref{fac} is a tree. Thus the BFE for \eqref{fac} is equal to the Gibbs free energy \cite[Theorem~3]{Yedida}, which is given by 
\begin{eqnarray}
{\rm G}(\{b_{k}, b_{N},\tilde b_{k}\})\triangleq-\sum_{k\in \mathcal K}\int b_{k}(x_k)\log b_{k}(x_k)dx_k \nonumber \\
-\int b_{N}(\matr x) \log\frac{p(\matr y\vert \matr x)}{b_{N}(\matr x)}d\matr x -\sum_{k \in \mathcal {K}}\int \tilde b_{k}(x_k)\log\frac{p_k(x_k)}{\tilde b_{k}(x_k)}dx_k\label{energy}. 
\end{eqnarray}
In this expression, $b_{N}$ and $\tilde b_{k}$, $k\in \mathcal {K}$, denote the beliefs of the factors, while $b_{k}$, $k\in \mathcal {K}$, denote the beliefs of the unknown variables in (\ref{fac}).

When we define a Lagrangian for \eqref{energy} that accounts for the set of marginalization consistency constrains, then at its stationary point, the belief $b_{k}(x_k)$ is equal to $p(x_k\vert\matr y)$ for all $k \in \mathcal {K}$ \cite{Yedida}. We consider the Gibbs energy formulation with a set of moment consistency constraints, instead of marginalization constraints. Specifically, following the arguments of \cite{Heskes} we define the Lagrangian
\begin{eqnarray}
&&\mathcal L(\{b_{k}, b_{N},\tilde b_{k}\})\triangleq{\rm G}(\{b_{k}, b_{N},\tilde b_{k}\})+ \mathcal Z \nonumber \\
&&-\sum_{k\in \mathcal K}\matr{\bar\nu}_k^\dagger\int \matr \phi(x_k)\left\{b_N(\matr x)-b_k(x_k)\right\}d\matr x\nonumber \\
&&-\sum_{k\in \mathcal K}\matr {\nu}_{k}^\dagger\int \matr \phi(x_k)\left\{\tilde b_{k}(x_k)-b_k(x_k)\right\}dx_k.
\label{lagrage}
\end{eqnarray}
The term $\mathcal Z$ accounts for the set of the normalization constraints for the beliefs:
\begin{eqnarray}
&&\mathcal Z\triangleq-\beta_N\left(1-\int b_{N}(\matr x)d\matr x\right) \nonumber \\
&&-\sum_{k\in \mathcal K} \beta_k \left(1-\int b_k(x_k)dx_k\right)-\tilde\beta_k\left(1-\int \tilde b_{k}(x_k)\right)\nonumber
\end{eqnarray}
with $\beta_N$, $\beta_k$, $\tilde\beta_k$, $k\in \mathcal {K}$ denoting the associated Lagrange multipliers. We consider constraints on the mean and variance, i.e. $\matr \phi(x_k)=(x_k,x_k^2)$. For convenience we write the Lagrangian multipliers explicitly appearing in \eqref{lagrage} in the form  
\begin{equation}
\matr{\nu}_k\triangleq\left[\gamma_k, -\frac{\lambda_k}{2} \right]^\dagger, \quad \matr{\bar\nu}_k\triangleq\left[\bar \gamma_k, -\frac{\bar\lambda_k}{2} \right]^\dagger, 
\quad k\in \mathcal {K}.
\end{equation}
We formulate the estimation procedure for $x_k, k\in \mathcal K$ as 
\begin{equation}
\mu_k\triangleq\int x_k b_k^{\star}(x_k)dx_k, 
\label{mmse}
\end{equation}
where $b_k^{\star}(x_k)$ represents the belief of $x_k$ at a stationary point of \eqref{lagrage}.
\subsection{\rm Stationary Points of the Lagrangian}\label{fixep}
In the sequel we derive the stationary points equations of the Lagrangian \eqref{lagrage}. For the sake of notational compactness we define \vspace{-0.1cm}
\begin{eqnarray}
&&\matr J\triangleq \frac{1}{\sigma_w^2}\matr A^\dagger \matr A, \qquad \matr \theta \triangleq \frac{1}{\sigma_w^2}\matr A^\dagger \matr y\\
&&\matr \Sigma \triangleq (\matr J+\matr {\bar\Lambda})^{-1} \label{def1}, \quad \matr \mu \triangleq\matr \Sigma (\matr\theta+\matr{\bar \gamma})\label{def}.
\end{eqnarray}
In \eqref{def} we have introduced the $K\times K$ diagonal matrix and the $K\times 1$ vector $\matr{\bar\gamma}$ whose entries are respectively $\bar \Lambda_{kk}=\bar\lambda_k$ and $\bar\gamma_k$, $k\in \mathcal K$. 

The stationary points of the Lagrangian \eqref{lagrage} are obtained to be of the form 
\begin{eqnarray}
\tilde b_{k}^{\star}(x_k)&=& \frac{1}{\tilde Z_k}p_k(x_k)\exp (\matr\nu_k^\dagger\matr\phi (x_k)),\quad k\in \mathcal K\label{belief}\\
b_{N}^{\star}(\matr x)&=& N(\matr x\vert \matr \mu, \matr \Sigma)\\
b_{k}^{\star}(x_k)&=& \frac{1}{Z_k}\exp ((\matr\nu_k+\matr{\bar\nu}_k)^\dagger\matr\phi (x_k)), \quad k\in \mathcal K.\label{belief1}
\end{eqnarray}
with ${\tilde Z_k}$ and ${Z_k}$ denoting the normalization constants for the beliefs in \eqref{belief} and \eqref{belief1}, respectively. At this stage it is convenient to define $\kappa_k\triangleq  \frac{\gamma_k}{\lambda_k}$, $k\in \mathcal K.$ With this definition we can rewrite the belief \eqref{belief} in the form
\begin{equation}
\tilde b_{k}^{\star}(x_k)= \frac{1}{Z(\kappa_k,\lambda_k)}p_k(x_k)N(x_k\vert \kappa_k,1/\lambda_k).\label{kappab}
\end{equation}
Furthermore we define for any $k\in \mathcal {K}$
\begin{eqnarray}
\eta(\kappa_k;\lambda_k)&\triangleq&\kappa_k+\frac{1}{\lambda_k}\frac{\partial \log Z(\kappa_k,\lambda_k)}{\partial \kappa_k},\\
\eta'(\kappa_k;\lambda_k)&\triangleq &\frac{\partial \eta(\kappa_k;\lambda_k)}{\partial \kappa_k}\label{etad}.
\end{eqnarray}
It is shown in  \cite[Eq.~(31)-(35)]{Florent} that $\eta(\kappa_k;\lambda_k)$ and $\eta'(\kappa_k;\lambda_k)/\lambda_k$ give the mean and the variance of the belief \eqref{kappab}, respectively. With these definitions, the identities resulting from the moment consistency constraints are given by
\begin{eqnarray}
b_{k}^{\star}(x_k)&=& N(x_k\vert \mu_{k},\Sigma_{kk})\quad k\in \mathcal K\label{b1}  \\ 
\frac{\lambda_k}{\eta'(\kappa_k;\lambda_k)}&=&\lambda_k+\bar\lambda_{k}, \quad k\in \mathcal K \label{f1} \\
\frac{\lambda_k\eta(\kappa_k;\lambda_k)}{\eta'(\kappa_k;\lambda_k)}&=&\gamma_k+\bar \gamma_k, \quad k\in \mathcal K\label{f2}.
\end{eqnarray}

We now derive a simple expression for \eqref{mmse}. By making use of the identities in \eqref{belief1} and \eqref{b1}, we write first
\begin{equation}
{\gamma}_k=\frac{{\mu}_k}{{\Sigma}_{kk}}- {\bar\gamma_k}, \quad  {\lambda_k}=\frac{1}{{\Sigma}_{kk}}-{\bar\lambda_k},\quad k\in \mathcal K\label{iden1}.
\end{equation}
Furthermore by the definitions in \eqref{def} we have 
\begin{equation}
\matr {\bar\gamma}=-\matr \theta+(\matr J+\matr {\bar\Lambda})\matr{\mu}.
\end{equation}
Let us introduce the $K\times K$ diagonal matrix $\matr {\Lambda}$ and the $K\times 1$ vector $\matr{\gamma}$ whose entries are respectively $\Lambda_{kk}=\lambda_k$ and $\gamma_k$, $k\in \mathcal K$. Then, making use of the identity in \eqref{iden1} we can write 
\begin{eqnarray}
\matr {\gamma}&=& \matr \theta- (\matr J+\matr{\bar\Lambda})\matr{\mu}+ \text{diag}(\matr\Sigma)^{-1}\matr{\mu} \\
&=& \matr \theta- \matr J\matr{\mu}+{\matr\Lambda}\matr{\mu}= \frac{1}{\sigma_w^2}\matr A^\dagger (\matr y-\matr A\matr{\mu})+{\matr\Lambda}\matr{\mu}
\end{eqnarray}
where $\text{diag}(\matr\Sigma)$ is the $K\times K$ diagonal matrix with $\text{diag}(\matr\Sigma)_{kk}=\Sigma_{kk}$, $k\in \mathcal K$. Then, by invoking the identities \eqref{b1} and\eqref{f2} we arrive at the sought explicit form for \eqref{mmse}:
\begin{eqnarray}
\mu_k  &=& \eta(\kappa_k;\lambda_k),\quad k\in \mathcal K \label{mu}\\
\kappa_k &=&\frac{1}{\lambda_k\sigma_w^2}\sum_{n\in \mathcal N}A_{nk}\left(y_n -\sum_{l\in \mathcal K}A_{nl}\mu_l \right)+\mu_k\label{fixes} \\
\lambda_{k} &=& \frac{1}{\Sigma_{kk}}-{\bar\lambda_{k}}, \quad  {\bar\lambda_{k}} = \frac{\lambda_k}{\eta'\left(\kappa_k;\lambda_k\right)}-\lambda_{k}. \label{fixe}
\end{eqnarray}
As a matter of fact equations \eqref{mu}--\eqref{fixe} coincide with the fixed point equations of ADATAP that are obtained by applying the \emph{cavity approach-new} \cite{Ole} in statistical physics, see \cite[Eq. (20), (25) and (26)]{Ole}. 

The step in \eqref{fixe} requires a matrix inversion, which is desirable to avoid in order to keep the complexity of fixed point algorithms devised from \eqref{mu}--\eqref{fixe} low. In \cite{Ole} the authors circumvent this complexity problem by using the so-called self-averaging method \cite[Section 3.1]{Ole} in the large system limit. The following theorem restates a result presented in \cite[Section 3.1]{Ole} in terms of the function $\eta'$ and the R-transform in free probability (see e.g. \cite{Ralf13}). \vspace{-0.3cm}
\begin{theorem}\cite[Section 3.1]{Ole}
Let $\matr A^\dagger \matr A$ have an AED as $N,K\to \infty$ with the ratio $\alpha= N/K$ fixed.
Let $\left<\eta'(\matr\kappa;\matr \Lambda)\right>\triangleq \frac{1}{K}\sum_{k\in \mathcal K}\eta'(\kappa_k;\lambda_k)$. Then, as $N,K\to \infty$ with the ratio $\alpha=N/K$ fixed, for all $k\in \mathcal K$ $\lambda_k$ converges almost surely to the macroscopic quantity $\lambda$ that is the solution of\footnote{By abusing the notation we define $\eta'(\matr\kappa;\lambda)\triangleq \eta'(\matr\kappa;\lambda\matr I)$, with $\matr I$ denoting the identity matrix of appropriate dimension.}
\begin{equation}
\lambda=\frac{1}{\sigma_w^2}{\rm R}_{\matr A} \left(-\frac{\left<\eta'(\matr\kappa;\lambda)\right>}{\sigma_{w}^2\lambda} \right) \label{perfect}
\end{equation}
with $\rm R_\matr A$ denoting the \emph R-transform of the AED of $\matr A^\dagger \matr A$.
\end{theorem}
Making use of the relation between the R-transform and the S-transform \cite[Table 6]{Ralf13} in \eqref{perfect} we obtain the following corollary.
\vspace{-0.3cm}
\begin{corollary}
Let the random matrix $\matr A$ be defined as in Theorem~1. Then, we have 
\begin{equation}
\lambda=\frac{1}{\sigma_w^2\rm S_{\matr A}\left(-\left<\eta'(\matr\kappa,\lambda)\right>\right)} \label{main}
\end{equation}
with $\rm S_{\matr A}$ denoting the \emph S-transform of the AED of $\matr A^\dagger \matr A$.
\end{corollary}
\section{Fixed Point Algorithms}
In this section we use the stationary point equations obtained in the previous section to introduce three fixed point iterative algorithms. Firstly we will present the classical EP scheme for \eqref{system} \cite{Seeger} and the ADATAP scheme \cite{Ole}. Secondly we derive the S-AMP algorithm mentioned in the introduction. 

All three recovery schemes have the following basis update step in common, which results by time-indexing the first identity in \eqref{def1}: 
\begin{equation}
\vspace{-0.1cm}
\matr \Sigma^t = (\matr J+\matr {\bar\Lambda}^t)^{-1}. \label{inv}
\end{equation}
Since only one element of $\matr {\bar\Lambda}^t$  is updated in each iteration the matrix inversion lemma can be applied to reduce the complexity of this step to $O(K^2)$, e.g. see \cite[Eq. (37)]{Ole5}. This makes \eqref{inv} suitable for applications with moderately large dimensions of $\matr A$.
\subsection{\rm EP and ADATAP}
In the following we present the compact form of the EP scheme for \eqref{system} (e.g. see \cite{Seeger}) and the ADATAP scheme \cite{Ole}. First we start with defining update steps  common to both algorithms. They follow by merely time indexing \eqref{fixe} for $k\in \mathcal K$:
\begin{eqnarray}
\lambda_{k}^t=\frac{1}{\Sigma^{t}_{kk}}-{\bar\lambda_{k}}^t, \quad {\bar\lambda_k}^t=\frac{\lambda_{k}^{t-1}}{\eta'(\kappa_k^{t-1};\lambda_k^{t-1})}-\lambda_{k}^{t-1}\label{lambdak}.
\end{eqnarray}

EP updates $\mu_k^t,k\in \mathcal K$ based on the second identity in \eqref{def}, \eqref{f1} and \eqref{fixes}:
\begin{eqnarray}
\mu_k^{t+1} &=& [\matr\Sigma^t (\matr\theta+\matr{\bar \gamma}^{t+1})]_k\label{REP}\\
\bar \gamma_k^{t+1}&=&  \frac{\lambda_k^t\eta(\kappa_k^{t};\lambda_k^t)}{\eta'(\kappa_k^{t};\lambda_k^t)}-\frac{\mu_k^t}{\Sigma_{kk}^t} \\
\kappa_k^t&=&\frac{1}{\lambda_k^t\sigma_w^2}\sum_{n\in \mathcal N}A_{nk}\left(y_n -\sum_{l\in \mathcal K}A_{nl}\mu_l^t \right)+\mu_k^t.  \label{kappa1}
\end{eqnarray}

ADATAP\cite{Ole} updates $\mu^{t}_k,k\in \mathcal K$ based on the stationary points identities in \eqref{mu}--\eqref{fixes}:
\begin{eqnarray}
\mu_k^{t+1}&=& \eta(\kappa_k^t;\lambda_k^t)\label {ADATAP}\\
\kappa_k^t&=&\frac{1}{\lambda_k^t\sigma_w^2}\sum_{n\in \mathcal N}A_{nk}\left(y_n -\sum_{l\in \mathcal K}A_{nl}\mu_l^{t}\right)+\mu_k^t. \label{kappa2}
\end{eqnarray}

Depending on the system model, EP and ADATAP may exhibit a poor convergence behavior, and may even diverge. A procedure to improve the convergence behavior consists in introducing a damping factor, say $\epsilon$, when updating e.g. $\mu^{t}_k$ in \eqref{kappa1} and \eqref{kappa2} as $(1-\epsilon)\mu_k^t+\epsilon \eta(\kappa_k^t;\lambda_k^t)\to \mu_k^t$. However this approach leads to very slow convergence which might require thousands of iterations, e.g. see \cite[Section V]{Kab}. Regarding more advanced damping schemes we refer the reader to \cite{Heskes2}.
\subsection{\rm S-AMP}
In the sequel we derive a new fixed point algorithm from the stationary points identities \eqref{mu}--\eqref{fixe}. The algorithm yields S-AMP in the large system limit. 

First we return to \eqref{fixes} and define 
\begin{equation}
z_{n,k} \triangleq \frac{1}{\lambda_k \sigma_w^2}\left(y_n -\sum_{l\in \mathcal K}A_{nl}\mu_l\right), \quad n\in \mathcal N, k\in\mathcal K\label{nice}.
\end{equation}
From this definition we ``devise" the following identity: 
\begin{equation}
z_{n,k}=y_n -\sum_{l\in \mathcal K}A_{nl}\mu_l+(1-\sigma_w^2 \lambda_k)z_{n,k}. \label{simple}
\end{equation}
Making use of  \eqref{mu}, \eqref{fixes} (with definition \eqref{nice}), and \eqref{simple} we obtain the new fixed point algorithm
\begin{eqnarray}
\mu_k^{t+1}&=& \eta\left(\sum_{n\in \mathcal N}A_{nk}z^t_{n,k}+\mu_k^t;\lambda_k^t\right), \quad k\in \mathcal K\label{SAMP}\\ 
z^t_{n,k}&=& y_n -\sum_{l\in \mathcal K}A_{nl}\mu_l^t+(1-\sigma_w^2 \lambda_k^{t-1})z_{n,k}^{t-1}\label{finite}
\end{eqnarray}
where $\lambda_k^t$ satisfies the system of equations
\begin{eqnarray}
\lambda_{k}^t= \frac{1}{\Sigma^{t}_{kk}}-{\bar\lambda_{k}}^t, \quad {\bar\lambda_k}^t=\frac{\lambda_{k}^{t}}{\eta'(\kappa_k^{t};\lambda_k^{t})}-\lambda_{k}^{t}\label{lambdak1}.
\end{eqnarray}
Like AMP, this scheme includes by design a natural damping factor $(1-\sigma_w^2 \lambda_k^{t-1})$ for the contribution $z_{n,k}^{t-1}$. Specifically in this scheme just like in AMP, we do not need a step-size parameter. However,at each iteration solving $\lambda_k^t$  from \eqref {lambdak1} is non-trivial in general. In this respect, the scheme in \eqref{lambdak} can be considered as an approximation of \eqref{lambdak1}.

By the design of $\lambda_k^t$ through \eqref{lambdak1}, and Theorem~1, for all $k\in \mathcal K$, $\lambda^t_k$ converges almost surely to a macroscopic quantity $\lambda^t$ as $N,K\to \infty$ with the ratio $\alpha$ fixed. Furthermore invoking Corollary~1 the quantity $\lambda^t$ is the solution of the identity  
\begin{equation}
\lambda^{t}=\frac{1}{\sigma_w^2{\rm S}_{\matr A}\left(-\left<\eta_{t}'(\matr\kappa^{t})\right>\right)} \label{fix1}
\end{equation}
where for convenience we define 
\begin{equation}
\eta_t(\kappa_k^t)\triangleq \eta(\kappa_k^t;\lambda^t) \quad k\in \mathcal K. \label{defeta}
\end{equation}
Consequently we obtain the iteration steps \eqref{keyb}-\eqref{key} of S-AMP in their scalar form:
\begin{eqnarray}
\mu_k^{t+1}&=&\eta_t\left(\sum_{n\in \mathcal N}A_{nk}z^t_{n}+\mu_k^t\right) ,\quad k\in \mathcal K \\
z^t_{n}&=& y_n -\sum_{l\in \mathcal K}A_{nl}\mu_l^{t} +\left(1-\frac{1}{{\rm s}_{\matr A}^{t-1}}\right)z_{n}^{t-1}.
\end{eqnarray}
We note that by the definition, $\matr \kappa^t=\matr A^\dagger\matr z^t+\matr \mu^t$.

In \cite{Bayati}, the function $\eta_t(\matr \kappa^t)$ in AMP updates is referred as ``an appropriate sequence of non-linear functions". By contrast, by design of the iterative process of S-AMP, we have the definition of $\eta_t(\matr\kappa^t)$ via the fixed point equation \eqref{fix1}. Note that, $\lambda^t$ must be solved at each iteration $t$ from this equation. Depending on the prior pdf's, obtaining closed form expression for $\lambda^t$ is often non-trivial. In fact this shows how S-AMP (or AMP in particular) can be a very advanced estimator as \eqref{fix1} directly relates the asymptotic stationary point identity in \eqref{main}. In order to better comprehend this aspect, in the following we examine $\lambda^t$ for the linear estimation problem.
\subsubsection{$\lambda^t$ for the Linear Estimation Problem}
The optimality of AMP for the linear estimation problem with the zero mean iid matrix ensemble, was proven in \cite[Section 2.1]{Bayati} via a minimization procedure upon the state evolution formula. We, by contrast, have the definition of S-AMP of which we can show the optimality for the general matrix ensembles.

Consider the linear observation model $\eqref{system}$. Let the entries of $\matr x$ be iid Gaussian with zero mean and variance one, i.e. $p_k(x_k)=N(x_k\vert 0,1)$, $k\in \mathcal K$. Then the asymptotic MMSE of \eqref{system} reads \cite{tulino}\footnote{In \cite{tulino}, the notation $\eta$ is used for \eqref{tau}. For convenience we adopt the notation $\tau$.}
\begin{equation}
\tau_{\matr A}(\sigma_w^2)\triangleq \int \frac{d{\rm P}_{\matr A}(x)}{1+\frac{x}{\sigma_w^2}} \label{tau}
\end{equation}
with ${\rm P}_{\matr A}(x)$ denoting the AED of $\matr A^\dagger \matr A$. Recall that the fixed points of S-AMP are the stationary points of the Gibbs free energy under the moment consistency constraints in the large system limit. Therefore, for the given a Gaussian prior, S-AMP must be a MMSE estimator in the large system limit. Namely the following relation must be satisfied:
\begin{equation}
\lim_{t\to \infty} \frac{\left<\eta_t'(\matr\kappa^{t})\right>}{\lambda^t}=\tau_{\matr A}(\sigma_w^2).\label{great}
\end{equation}
We show next that actually \emph{for any $t\geq 0$}, $\left<\eta_t'(\matr\kappa^{t})\right>/\lambda^t=\tau_{\matr A}(\sigma_w^2)$. First notice that with the choice of the prior we have $\left<\eta_t'(\matr\kappa^{t})\right>=\lambda^t/(1+\lambda^t)$. From the definition in \eqref{fix1}, we have 
\begin{equation}
\lambda^t=\frac{1}{\sigma_w^2{\rm S}_{\matr A} \left(-\frac{\lambda^t}{1+\lambda^t}\right)}.
\end{equation}
The S-transform can be formulated in terms of $\tau_{\matr A}(\sigma_w^2)$ \cite[Definition 2.15]{tulino}. Using this formula we write
\begin{equation}
\frac{1-\tau_{\matr A}(\sigma_w^2)}{\tau_{\matr A}(\sigma_w^2)}=\frac{1}{\sigma_w^2{\rm S}_{\matr A} \left(\tau_{\matr A}(\sigma_w^2)-1\right)}.
\end{equation}
Thus $\lambda^t= 1/\tau_{\matr A}(\sigma_w^2)-1$, which confirms the optimality of S-AMP for the linear estimation problem.
\section{A Sub-Optimal Variant of S-AMP}\label{SubOptSc}
In the previous subsection we derived the explicit expression for $\lambda^t$ when the prior pdf's are Gaussian. However solving $\lambda^t$ from \eqref{fix1} for the other prior pdf's is often non-trivial. A direct approach consists in including an inner loop to solve \eqref{fix1} iteratively at each iteration. That would, however, create an overhead that we would like to avoid. Instead, we propose in the following a sub-optimal scheme for $\lambda_t$ that does not require any inner loop. We approximate the optimal $\lambda^t$ defined by \eqref{fix1} with $\lambda_s^t$ that satisfies 
\begin{eqnarray}
\lambda_s^{t} &=&\frac{1}{\sigma_w^2{\rm S}_{\matr A}\left(-\frac{\lambda_s^{t}}{\lambda_s^{t-1}}\left<\eta_{t-1}'(\matr\kappa^{t-1})\right>\right)}\label{suboptimal}.
\end{eqnarray}
Here we note that, the sub-optimal scheme coincides with the same fixed point equations of the optimal scheme. 

When the entries of $\matr A$ are iid with zero mean and variance $1/N$, the sub-optimal scheme coincides with the classical recursion of AMP used in the literature e.g. \cite{Florent}. In fact, from \eqref{suboptimal},  it is easy to obtain the so-called state-evolution formula\cite{Donoha1} for the iid zero mean matrix ensemble. Finally, we note that it is possible to introduce more advanced recovery schemes. But this is out of the scope of this contribution. 

In the sequel we assess the performance of the sub-optimal variant of S-AMP. Due to the space limitation we only consider the system model used in \cite[Section 5]{Kab} for Bayesian inference in compressed sensing. Accordingly, the prior pdf's are Bernoulli-Gaussian: $p_k(x_k)=(1-\rho)\delta(x_k) +\rho N(x_k\vert 0, 1)$, $k\in \mathcal K$. We refer the reader to \cite[Eq.~(67) \& (68)]{Florent} for the closed form expressions of $\eta_t(\cdot)$ and $\eta_t'(\cdot)$ in this case. 

We consider the sub-optimal variant of S-AMP for two scenarios: i) the random row-orthogonal matrix ensemble, i.e.\  $\matr A=\alpha^{-\frac{1}{2}}\matr P_{\alpha} \matr O, \alpha\leq 1$, where $\matr P_{\alpha}$ is the $N\times K$ matrix with entries $[\matr  P_\alpha]_{ij} = \delta_{ij}, \forall ij$, with $\delta_{ij}$ denoting the Kronecker delta, and $\matr O$ is the $K\times K$ Haar matrix \cite{Mikko}; ii) iid zero mean Gaussian matrix ensemble. Note that in the latter case, the sub-optimal variant coincides with the classical AMP recursion as in \cite{Florent}. In the former case, with a straightforward calculus in free probability we obtain that ${\rm S}_{\matr A}(\omega)=(1+\omega)/(1+\omega/\alpha)$ and 
\begin{equation}
\lambda_s^{t}=(1+\chi^{t}-\sqrt{(1+\chi^{t})^2-4\alpha\chi^{t}})/(2\alpha\sigma_w^2\chi^{t})\label{key2}
\end{equation}
where $\chi^t\triangleq \left<\eta_{t-1}'(\matr A^\dagger\matr z^{t-1}+\matr\mu^{t-1})\right>/(\alpha\sigma_w^2 \lambda_s^{t-1})$. 
\begin{figure}
\begin{center}
\epsfig{file=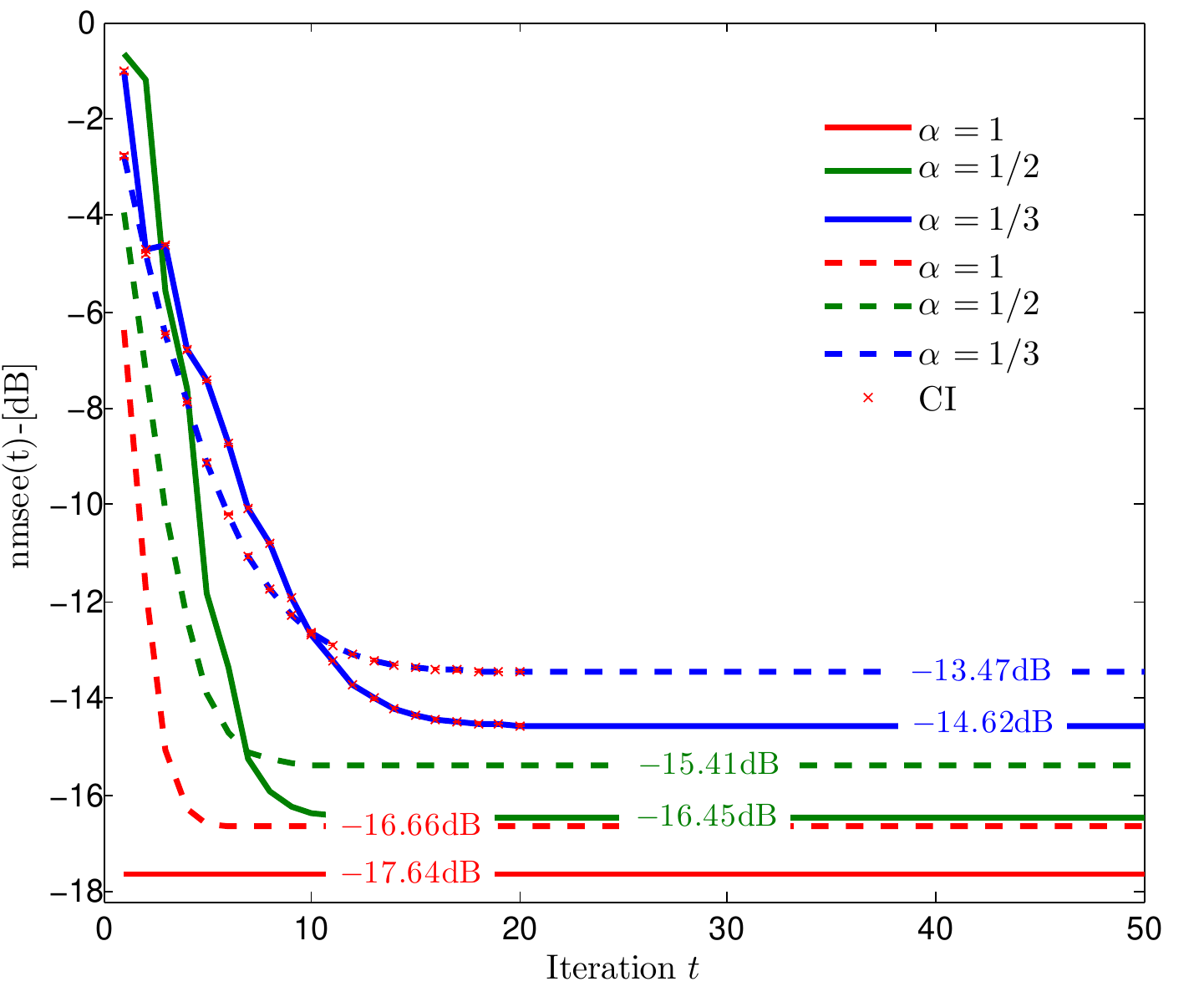,width=6cm}
\vspace{-0.2cm}
\caption{Performance of the sub-optimal variant of S-AMP implemented for the row orthogonal matrix ensemble (solid curves) and the iid zero mean Gaussian matrix ensemble (dashed curves). Note that in the latter case the scheme leads to the classical AMP recursion. The empirical nmsee per iteration is reported versus the number iterations for different selections of $\alpha$. Confidence intervals (CIs) are also shown for $\alpha= 1/3$. We set $\sigma^2_w=-20$~dB and $\rho = 0.1$. For each selection of $\alpha$ 2000 trials are performed.} 
\end{center}
\vspace{-0.5cm}
\end{figure}

In \cite[Section 5]{Kab}, the authors report the estimated normalized 
mean square estimation error (nmsee) of the damped-ADATAP scheme for the setting (i). For each 
trial up to 3000 iterations are executed. 
In Figure 1 we report the nmsee of the suboptimal variant of S-AMP applied in the same context versus the number of iterations. Details are reported in the caption of Fig.~1. Note that no divergence behavior was observed in all performed trials. A comparison of the 
curves in Fig. 1 with the corresponding curves reported in \cite[Fig.~1]{Kab} 
show that both recovery schemes achieve the same performance, but with a 
significantly smaller number of iterations for the sub-optimal variant. 
\section{Conclusion}
We developed a novel low-complexity fixed-point algorithm for linear observation systems from the equations of the stationary points of the exact Gibbs free energy under first- and second-moment consistency constraints in the large system limit. The algorithm that we call S-AMP extends AMP for general matrix ensembles. Specifically, AMP is a special case of S-AMP when the measurement matrix has iid zero mean entries. The optimality of S-AMP follows by its design. Furthermore, we define a sub-optimal variant of S-AMP, which is easy to implement. This sub-optimal recovery scheme shows excellent performance for the row-orthogonal matrix ensemble in compressed sensing and it converges in around 40 iterations without showing any divergence behavior.
\bibliographystyle{IEEEtran}
\bibliography{IEEEabrv,liter}
\end{document}